# Avoiding the Detector Blinding Attack on Quantum Cryptography

Z L Yuan, J F Dynes and A J Shields

Toshiba Research Europe Ltd, Cambridge Research Laboratory, 208 Cambridge Science Park, Milton Road, Cambridge, CB4 0GZ, UK

Although the protocols used for quantum key distribution (QKD) have been proven to be unconditionally secret, the security of actual hardware depends critically on the detail of their implementation. Studies of quantum hacking play an important role in exposing potential weaknesses and thereby promote the design of more secure systems. In their recent paper[1], Lydersen *et al.* presented an attack on QKD systems that would – allegedly – allow an eavesdropper to gain full information about the secret key. The authors state that "the loophole is likely to be present in most QKD systems using avalanche photodiodes to detect single photons". Here, we show the attack will be ineffective on most single photon avalanche photodiodes (APDs) and certainly ineffective on any detectors that are operated correctly. The attack is only successful if a redundant resistor is included in series with the APD, or if the detector discrimination levels are set inappropriately.

Lyderson's attack targets the InGaAs/InP APDs often used to detect single photons. Figure 1a shows a typical biasing circuit for gated Geiger mode APDs, unusual only in the inclusion of a redundant biasing resistor ($R_{bias}$) used here to simulate Lydersen's experiment. The APD is pulse-gated to raise its bias ($V_g$) above the breakdown voltage ($V_b$). When biased above breakdown ($V_g > V_b$), the device can multiply a single-photon induced charge into a macroscopic current through repeated impact ionisation. A detection event is registered if the voltage drop across the sensing resistor ($R_s$) exceeds the discrimination voltage level ($L$). It is good (and common) practice to set this discrimination level as low as possible,[2-4] determined in gated mode by the capacitive charging signal ($L_0$ in Fig.1a, inset).

Lyderson *et al* send strong CW illumination along the fibre to generate a photocurrent-induced voltage drop across the bias resistor $R_{bias}$, thereby reducing the APD bias below the breakdown voltage (*ie.* $V_g < V_b$) and rendering the detector blind to single photons[5]. As demonstrated in Fig. 1b, we find the APD is indeed blind (*i.e.* the count probability falls to zero) at a threshold CW power of 22nW, 350nW, and 2.4μW for values of $R_{bias}$ of 680k, 330k and 100kΩ, respectively. We find the count probability recovers to one count per gate under stronger illumination at around 20 μW, irrespective of the value of $R_{bias}$. The detector triggers under strong illumination (>20µW) due to modulation of the photocurrent gain by the applied bias pulses (see Fig. 1a inset). As described below, this gain modulation negates the detector blinding attack, provided that the discrimination level is set appropriately.

Notice in Fig.1b that the range of CW input powers over which the detector is blind to single photons narrows as $R_{bias}$ decreases. Indeed for sufficiently small $R_{bias}$, the detector cannot be blinded. This is exactly what we observe for the usual case of $R_{bias}$=0 (Fig. 1b). We stress that although a biasing resistor is sometimes used for quenching avalanches in APDs operated in dc mode, it is redundant for gated Geiger mode and not common. Thus most APDs used in QKD will not be sensitive to the detector blinding attack[6], contrary to the suggestion by Lydersen et al.[1]

For finite biasing resistors, as is the case for the QPN5505 and Clavis2 QKD systems studied by Lydersen et al, the ability to blind the detector with CW input light is very sensitive to the detector discrimination level. This is illustrated by Fig.1c which shows the CW power dependence of the count probability for $R_{bias}$=1kΩ, the value appropriate to Clavis2, and two different discrimination levels. Notice if the discrimination level ($L$) is set just above the capacitive signal, i.e., $L=L_0$, the detector cannot be blinded. However, if the discriminator is set to an inappropriate level ($L=2L_0$), the detector is blinded at 260 µW, close to the value reported by Lydersen et al. A later report by the same authors[7] reports setting a discrimination level of about 80mV for Clavis2, which is more than twice the value needed to reject the capacitive signal of 35mV. This suggests that setting appropriate detector discrimination levels would be sufficient to prevent the detector blinding attack on the QPN5505 and Clavis2 QKD systems also.

We point out that for appropriate discrimination levels, gain modulation of the photocurrent will also be sufficient to prevent the thermal attack[7] on APDs. This attack uses higher CW powers to generate a photocurrent sufficient to heat the APD and thereby increase $V_b$. Considering the required optical power (>1mW)[7], a slight modulation in gain is sufficient for persistent counting. Indeed we have confirmed the absence of any thermal blinding effect with an optical excitation of 17.8 mW, corresponding to a heating power of 500 mW in the APD.

Finally, we emphasize that any attack with strong illumination will result in a large photocurrent. Monitoring the photocurrent for anomalously high values is a straightforward way to detect any attack of this type. This method is applicable to all types of APDs, including those which are un-gated[5] or used in high-speed gated mode[3,4], thereby revealing bright illumination attacks on QKD systems using APDs.

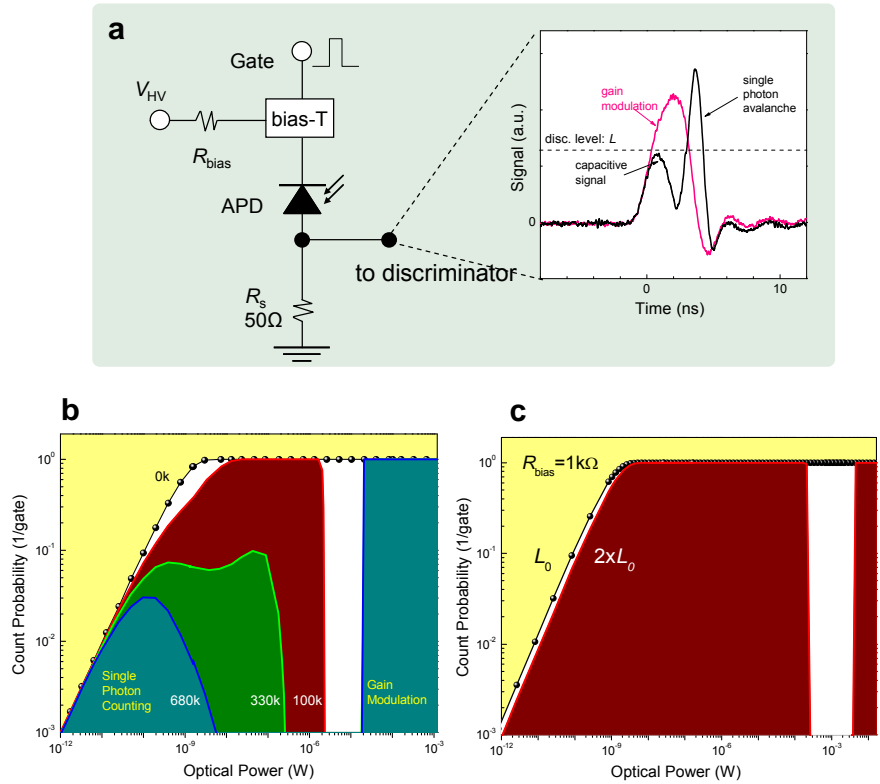

**Figure 1** InGaAs APD under CW illumination. The APD is cooled thermal-electrically with a setting temperature of -30 °C and driven by voltages pulses of 3.5 ns, 4 V and 2 MHz. The excess bias is set at 2.5 V by the DC bias in the single photon counting regime. A 1.55 μm CW laser is used for illumination. **a**, A schematic diagram for gated mode operation and two APD output waveforms that trigger a photon click. **b**, count probability *vs.* incident optical power for a biasing resistor ($R_{bias}$) of differing impedance. **c**, count probability vs. incident optical power for two differing discrimination levels: $L_0$ and $2 \times L_0$.